# Recyclable vitrimer-based printed circuit board
# for circular electronics

Zhihan Zhang[1], Agni K. Biswal[2], Ankush Nandi[2], Kali Frost[3], Jake A. Smith[1,3],

Bichlien H. Nguyen[1,3], Shwetak Patel[1], Aniruddh Vashisth[2*], Vikram Iyer[1*]

[1] Paul G. Allen School of Computer Science & Engineering, University of Washington

[2] Department of Mechanical Engineering, University of Washington

[3] Microsoft Research

* Corresponding authors: vashisth@uw.edu or vsiyer@uw.edu

# Abstract

Electronics are integral to modern life; however, at their end-of-life these devices produce environmentally hazardous electronic waste (e-waste). Recycling the ubiquitous printed circuit boards (PCBs) that make up a substantial mass and volume fraction of e-waste is challenging due to their use of irreversibly cured thermoset epoxies. We present a PCB formulation using transesterification vitrimers (vPCBs), and an end-to-end fabrication process compatible with standard manufacturing ecosystems. We create functional prototypes of IoT devices transmitting 2.4 GHz radio signals on vPCBs with electrical and mechanical properties meeting industry standards. Fractures and holes in vPCBs can be repaired while retaining comparable performance over more than four repair cycles. We further demonstrate non-destructive decomposition of transesterification vitrimer composites with solid inclusions and metal attachments by polymer swelling with small molecule solvents. We hypothesize that unlike traditional solvolysis recycling, swelling does not degrade the materials. Through dynamic mechanical analysis we find negligible catalyst loss, minimal changes in storage modulus, and equivalent polymer backbone composition across multiple recycling cycles. We achieve 98% polymer recovery, 100% fiber recovery, and 91% solvent recovery which we reuse to create new vPCBs without degraded performance. Our cradle-to-cradle life-cycle assessment shows substantial environmental impact reduction over conventional PCBs in 11 categories.

# Main

Electronics have become an increasingly integral part of modern life in everything from consumer devices like smartphones and laptops to state-of-the-art industrial, scientific and medical technologies. However, when electronics reach the end of their life cycle due to obsolescence or damage, they become electronic waste (e-waste) and pose significant environmental hazards. E-waste contains a complex toxic mixture of various metals, silicon integrated circuits (ICs), glass fibers, thermoset polymers, flame retardants, and more, which can pollute the air, soil, and water posing significant hazards for the surrounding communities[1–3]. With over 53.6 million metric tons (Mt) generated in 2019 alone[4], e-waste is one of the fastest-growing waste streams globally[5] and a matter of global concern[6].

In response to this issue, efforts are underway to reduce e-waste by transitioning to a circular economy model in which electronics can be transformed into secondary raw materials. A key challenge in achieving a circular manufacturing cycle is the recycling of printed circuit boards (PCBs). PCBs are ubiquitous in electronics and rank within the world's top 100 most traded products[7]. They form the physical substrate upon which chips are mounted and connected with metal traces patterned on their surface. PCBs are most commonly composite materials made of glass fiber weaves within a flame-retardant thermoset epoxy matrix (FR-4). The majority of prior recycling efforts have focused on the recovery of intact ICs and high-value minerals such as gold and copper[8,9]; however 70% of a PCB's volume and mass is made up of dielectric substrates[10].

To maximize sustainability, we aim to create a PCB that can be repeatedly recycled to produce brand-new circuits with the same performance. Achieving a scalable and closed-loop circular manufacturing cycle presents several challenges. First, PCBs must be capable of tolerating electronics manufacturing processes which include chemical etching, electroplating, and elevated-temperatures for soldering. Prior water-soluble[11–16] and healable materials[17] sacrifice compatibility with these methods to achieve recyclability which limits their potential to scale. Additionally, these materials do not meet the rigorous standards for moisture absorption, flammability, dielectric constant, loss tangent, and other criteria needed to produce high-speed and RF circuits common in modern electronics. Second, the irreversibly cured thermosets in the dielectric substrates like FR-4 are formulated to be resistant to fire and chemicals, which makes it incredibly challenging to fully deconstruct them into raw materials for recycling. While substituting thermoplastics in place of these thermosets is a potential solution, thermoplastic reprocessing damages the polymer chains[18], which reduces their thermal and mechanical properties and prevents their reuse in new PCBs. Third, closed-loop recycling imposes the strict constraint of deconstructing composites into raw materials *without damage*. A number of methods have been explored to recycle these materials however they all result in substantial material damage[19,20]. Direct mechanical grinding and crushing of waste PCBs are among the simplest, however it is incompatible with recycling composite materials like FR-4. This process destroys the glass fiber and polymer, preventing them from being reused for making a new board. Mechanical recycling is used to homogenize the PCB so that it can be put into a furnace or subjected to a chemical leaching process for commercial metal recovery, however, these methods also suffer from low efficiency in the separation of metals. In contrast, thermal decomposition (pyrolysis), or chemical dissolution in strong solvents (solvolysis) have higher efficiency but require high processing temperatures which increases energy costs[21], or require nonreusable solvents[21,22] thereby producing hazardous byproducts[23]. Additionally, many of these methods are only able to recycle the glass fiber and dissolve the polymer chains completely[22]. Because thermosets lack the ability to reform bonds, the material properties of the recycling outputs are degraded thereby limiting their application potential and reuse cycles.

In this article, we present a different solution by introducing a novel PCB formulation using transesterification vitrimers (vPCBs) that are specifically designed for circularity. We then develop techniques for vPCB repair, and non-destructive recycling methods for vitrimer composites with solid inclusions by polymer swelling using small molecule solvents accompanied by analyzing the underlying mechanisms. Recent work has shown that dynamic covalent adaptive networks (CANs) with associative exchange reactions can undergo reversible crosslinking at elevated temperatures[24]. These materials, termed vitrimers, exhibit dynamic covalent bond exchange reactions when heated. This enables the unique healing capabilities of vitrimers[25] and repeated reversal of fatigue-induced damage to recover properties close to those of the original material[26], allowing the recovered materials to be reused repeatedly with minimal degradation. We specifically focus on vitrimers with associative exchange reactions over dissociative mechanisms due to their ability to maintain a high crosslink density over multiple recycling cycles. Replacing the traditional thermoset epoxy matrix with a vitrimer addresses the fundamental recycling challenge described above. vPCBs not only enable the repair of PCBs but also open up the potential for high-value recycling by reusing the polymer to create a new, high-performance PCB. Further, these methods could be generalized to this entire class of vitrimer polymers whose backbones can be tuned to achieve desired material properties.

Fig. 1 illustrates our fully circular manufacturing and recycling pipeline which highlights our key contributions. First, we create glass fiber-reinforced vitrimer composites and develop an end-to-end PCB fabrication process compatible with current electronic manufacturing services (EMS) ecosystems including multi-layer copper lamination, chemical etching, laser structuring, electroless plating, and soldering. We perform detailed characterization of our vPCBs demonstrating they meet industry standards for peel strength of copper-clad laminates, flexural strength, moisture absorption, dielectric constant, and resistivity. We then use them to demonstrate a fully functional IoT sensor capable of transmitting 2.4 GHz Bluetooth signals. Second, we leverage the bond exchange capabilities of vitrimers to repair physical damage to vPCBs. We repair fractures, holes and mechanical deformations, demonstrating that repaired vPCBs maintain their mechanical and electrical properties over more than four repair cycles. We verify the absence of microcracks and pores using scanning electron microscopy (SEM) and show that covalent bonding achieves higher joint strength than cyanoacrylate adhesives in shear punch tests.

Third, we develop an end-to-end recycling process for our vPCBs. We observe that certain solvents can cause the vitrimer to swell, and leverage this polymer swelling in small-molecule solvents to deconstruct transesterification vitrimer composites with solid inclusions and metal attachments. We hypothesize that unlike traditional solvolysis recycling, swelling does not degrade the materials; however, this method raises a number of fundamental questions of whether the process removes non-covalently bonded catalyst critical to the transesterification vitrimer's properties, effects on the crosslink density, and integrity of the polymer backbone. We develop a process to create recycled samples and comprehensively investigate these questions across multiple recycling cycles. We perform dynamic mechanical analysis revealing negligible change in vitrimer transition temperature ($T_v$) of $\pm 0.7$ °C indicating no catalyst loss, and minimal change in storage modulus (3.1%) which depends on crosslink density. We confirm with Fourier-transform infrared spectroscopy (FTIR) spectroscopy that the molecular structure remains unchanged, and verify recycled vPCBs have mechanical and electrical properties on par with the original. Our recycling process achieves 98% recovery of the vitrimer polymer and 100% recovery of glass fibers, and even a 91% recovery of the solvent which we successfully reuse. Fourth, we perform a comprehensive cradle-to-cradle life-cycle assessment (LCA) leveraging industrial scale models to compare the environmental impact of our remanufacturing and recycling systems with conventional PCBs. Our results show a 47.9% improvement in global warming potential, 65.5% in ozone depletion potential, and 80.9% in human cancer toxicity emissions for our recycling process compared to conventional PCB end-of-life disposal scenarios.

**Fig 1. Transesterification vitrimer-based fully recyclable PCB**

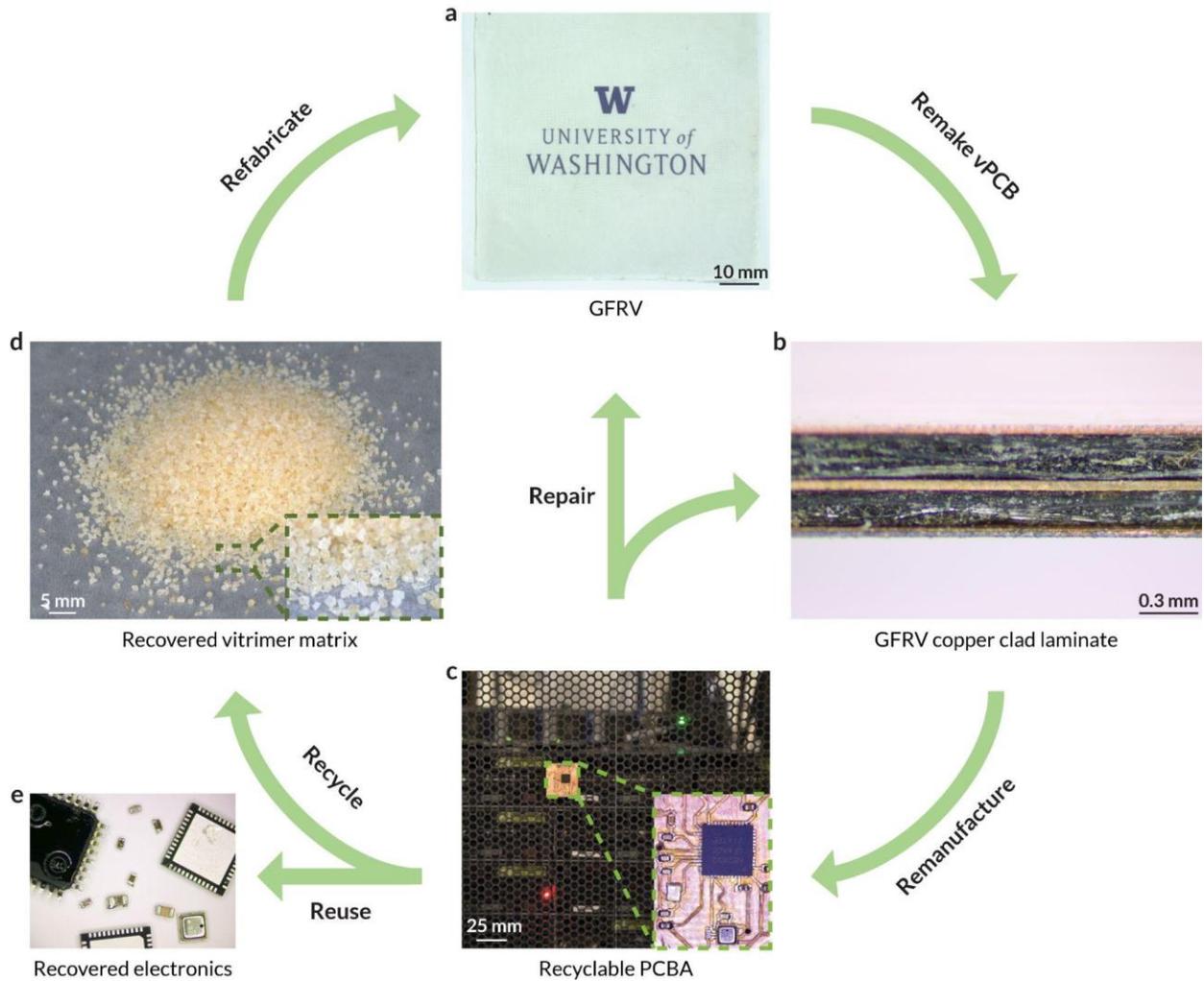

Conceptual diagram showing the closed-loop repair and recycling of vPCB. **a,** GFRV composite. **b,** Cross-sectional view of a three-layer vPCB. **c,** Photograph of a vPCB-based recyclable IoT next to a server. The printed circuit board assembly (PCBA) can be repaired if warping or damage occurs in the base material. After the device is decommissioned, the substrate materials can be recycled and the electronic components could be reused. **d,** Recovered vitrimer from obsolete vPCB after drying and pulverizing, which can be reused to fabricate new GFRVs.

## Results

### Glass fiber reinforced vitrimer composites

To enable PCB circularity, we seek to disassemble the entire board into its constituent raw materials without damage for reuse. While the electronics and copper traces can be recovered using thermal exposure[27] or the same chemical etchants used to pattern the circuits[28], the dielectric substrate material remains. The most common material, FR-4 is a combination of a woven glass fiber impregnated with a thermoset epoxy resin. The covalent crosslinks of these thermosets give the material desirable properties such as high structural integrity, chemical stability, and resistance to temperature. These properties, however, present a trade-off common in sustainability: the material's robustness makes it very difficult to effectively separate the epoxy and glass fibers for recycling or reuse.

In this work, we take a different approach inspired by recent advancements in healable materials to re-engineer PCBs using polymers with dynamic covalent adaptable networks (CANs); vitrimers are a special class of CANs with a thermosetting macromolecular network with healing abilities. This property also enables entirely new capabilities. First, it enables remanufacturing, in which holes in the material can be refilled and segments can be tiled together without transforming them back into raw starting materials. Second, the ability to reuse vitrimers as secondary raw materials opens the possibility for effective recycling if they can be non-destructively separated from the glass fibers. Our vitrimer polymers are synthesized from a bifunctional epoxide (DGEBA) and an acid (adipic acid) in the presence of a catalyst Triazabicyclodecene (TBD). The chemical structures for the reaction are shown in Fig. 2a. In contrast to the prior attempts at integrating healable polymers into circuits[29], our vitrimer chemistry is optimized for similarity to conventional FR-4 by choosing a bisphenol-A epoxide which is typically used in PCBs[30]. The normalized stress-relaxations at temperatures ranging from 140 °C to 240 °C and the characterized storage modulus and tan delta for vitrimer with a 5 mol % catalyst concentration are shown in Extended Data Fig. 1. It is noteworthy that by tuning the specific components and functional groups (Fig. 2d), a wide variety of different properties can be achieved[31–33], such as a higher glass transition temperature (Tg) of 146 °C[34].

Our recyclable GFRV composites are created using a scalable fabrication process similar to conventional FR-4 fabrication shown in Fig. 2b (see Methods for additional details). Figure 2c schematically illustrates the structure of the resulting GFRV.

Moisture absorption of GFRV is characterized through immersion in water according to the IPC PCB standard (see Methods for additional details). Moisture can accelerate various failure mechanisms in PCBs by causing cracking, delamination, soldering issues, and changes in dielectric properties. High moisture absorption is a common failure mode for many sustainable PCB materials, as moisture ingress usually weakens polymer by hydrolytic cleavage in chains[35]. Fig. 2e shows our vitrimer is on par with common PCB dielectrics like polyimide and within 0.2% of the FR-4 standard. We hypothesize this difference is due to the presence of the ester linkages between monomers, and this value could be further reduced by modifying the vitrimer chemistry.

Flexural strength is also evaluated using a three-point flexural test (see Methods) and the results are visualized in Fig. 2f. The performance is within the range of FR4-based materials and these values could be tuned by changing the quality of the glass fiber weave.

**Fig. 2: Glass fiber reinforced vitrimer composite.**

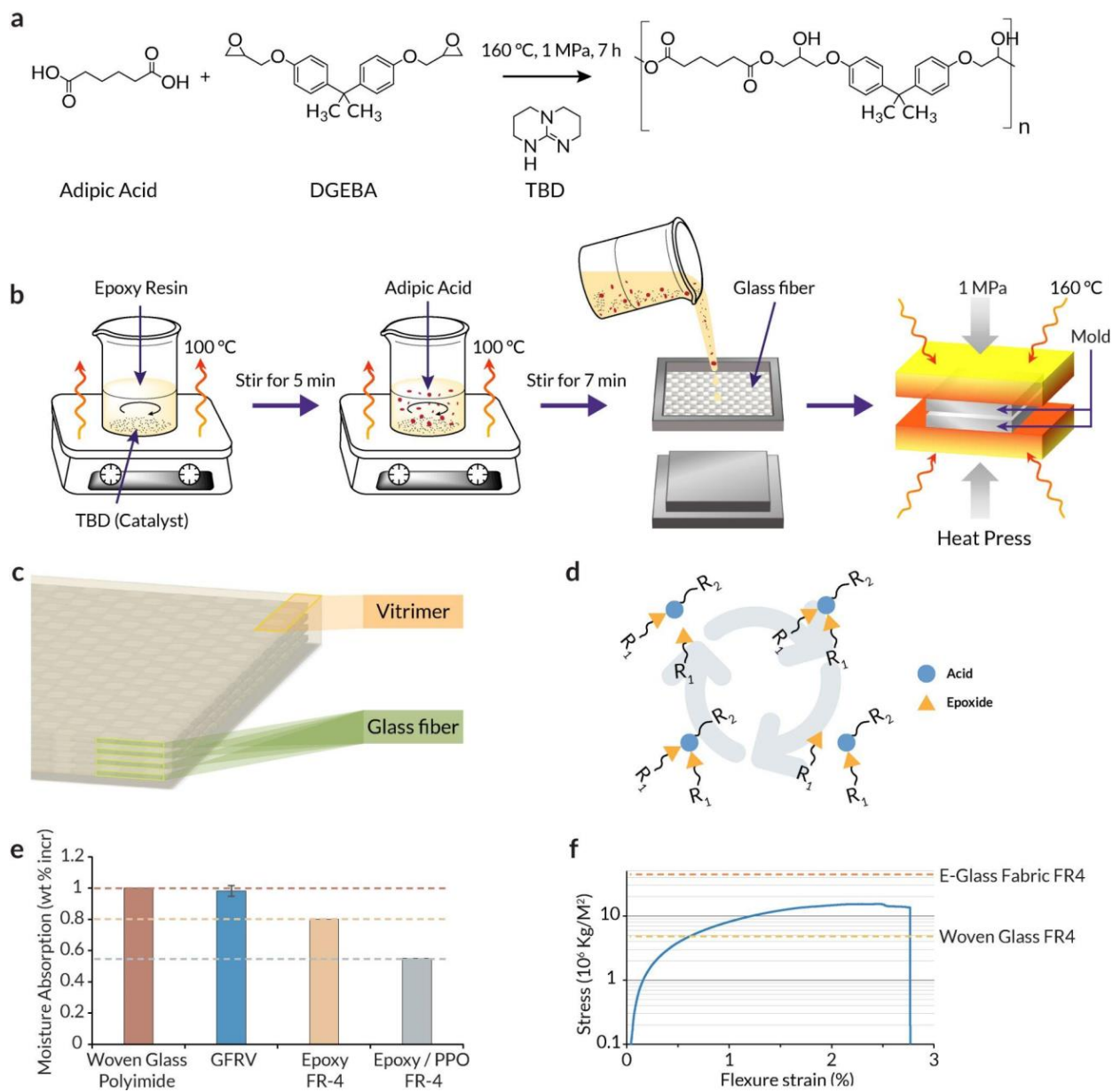

**a,** Reaction of transesterification vitrimer under heat and pressure. **b,** Schematic of the fabrication process of GFRV. **c,** Schematic of four-layer glass fiber GFRV. **d,** Bond exchange via transesterification in vitrimer networks enables healing and recyclability when heated. Properties can be tuned by changing the functional chains $R_1$ and $R_2$ in the epoxide and acid used. **e,** Characterized moisture absorption of GFRV compared to the PCB standards ($N = 3$, error bar = $\pm\sigma$). **f,** Characterized flexural strength of GFRV compared to the PCB standards of woven glass FR-4 and woven e-glass fabric FR-4.

**Vitrimer-based PCB**

We create *recyclable*, *reusable*, and *repairable* circuit boards by integrating these unique vitrimer materials into PCBs. The first step to creating a PCB from the raw GFRV is to add a conductive copper layer for circuit traces. In traditional PCB manufacturing, a glass fiber weave is impregnated with a phenolic epoxy resin, and partially cured. The resulting material, known as prepreg, is then laminated with thin copper foil and can be patterned and stacked to create multilayer circuits. To create the vPCBs, a similar lamination process is used as illustrated in Fig. 3a. Sheets of copper foil are laminated on the raw GFRV in a heat press (see Methods for details). This method can easily scale to roll-to-roll manufacturing for high-volume production.

To assess the quality of vPCB copper adhesion, the peel strength for 16 copper-clad laminates (CCLs) is systematically measured for heat press temperatures of 120-160 °C and heat press time of 30 min-2 h. Empirical measurements were performed due to the complexities of accurately modeling the adhesion bonds formed by heat and pressure on polymer surfaces. Our results show that the interfacial adhesion between copper and vitrimer increases as the heat press temperature and time increase (Fig. 3b). We note however that above 180 °C lower viscosity causes the vitrimer to be squeezed out. The results follow the expected trend, whereby the increase in temperature and pressure results in dissociation in the covalent adaptive network, thereby increasing the availability of functional reactive sites for adhesion. Bonding can be further improved by adding a thin, partially cured vitrimer layer between GFRV and copper foil This significantly enhances peel strength after thermal stress, exceeding the PCB standard (Extended Data Fig. 2), as the partially cured vitrimer has more reactive sites available.

Next, circuit traces are patterned onto the raw copper-coated substrate. Chemical etching and laser structuring are the two commonly used processes in PCB manufacturing. Laser structuring is often used for more intricate circuits due to its high precision, ability to cut through a wider range of materials, and speed in small-scale manufacturing. Chemical etching is less precise but is more suited to low-cost, large-scale production because it can etch many parts or multiple boards simultaneously. We demonstrate compatibility with both manufacturing processes. Briefly, the raw copper-coated GFRV laminate is placed in a laser micromachining system (LPKF Protolaser U4), where the laser is used to both drill holes and directly remove copper to create a circuit. Alternatively, we can coat the copper with a thin mask layer and use the laser to remove the mask in selective regions, then place the board in ferric chloride solution to etch away the exposed copper. This method can be used to pattern individual layers of material that can then be aligned and heat pressed to create a multi-layer laminate. The final step is applying an additional mask layer and using an electroless copper plating process to connect the layers through vias (See Methods for additional details).

The fabricated circuits are characterized to compare vPCBs to traditional FR4-based circuits. The dielectric constant of the PCB substrate material is crucial for achieving good high-frequency electrical performance, as it affects the intensity of signal reflections and crosstalk, which ultimately impacts the overall signal integrity. The measured dielectric constant and loss tangent of the vPCB substrate compared to the PCB standards of various FR-4 types are shown in Fig. 3c, d, respectively (details on characterization procedures are provided in the Methods). The results show that the dielectric constant for our vPCB is within the typical range of 3.5 to 5.5, demonstrating it meets the required specifications. We note that this value can also be tuned to meet the needs of the application by changing the glass fiber volume content, vitrimer chemistry, and polymer-free volume.

Low flammability is another crucial property required to protect PCBs from potential ignition as many electronic components produce heat and contain flammable polymers. Conventional PCB substrates such as FR-4 are formulated with flame retardants to self-extinguish potential fires. To evaluate the flammability of vPCB, a GFRV is rigorously tested using a burner, and found to exhibit excellent fire-retardant properties. As shown in Fig. 3e, the fire was extinguished within 7 seconds after the removal of the flame source (Fig. 3e, Supplementary Video 2), conforming to the 10-second standard.

**Fig. 3: Characterization of vPCB.**

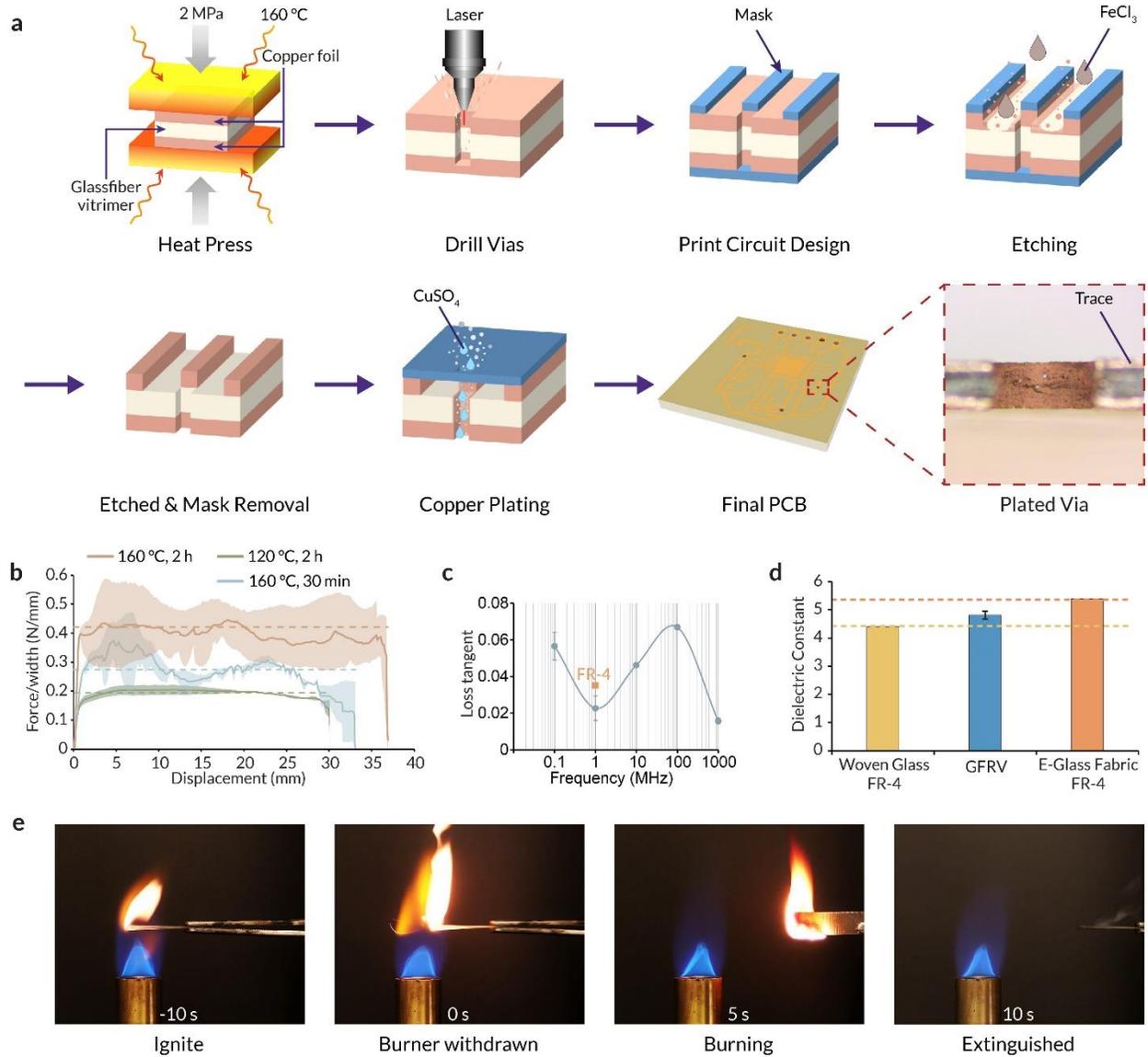

**a,** Schematic of the scalable manufacturing process of vPCB. **b,** Curves of the peeling force per width of copper-clad versus displacement for laminates with various heat press temperatures and times ($N \geq 3$, shaded region indicates max and min values of trials). **c, d,** Characterized dielectric constant (**c**), and loss tangent (**d**) of vPCB compared to the PCB standards of various FR-4 ($N = 3$, error bar = $\pm\sigma$). **e,** Flammability test of vPCB. The sample is retracted after 0 s to move it away from the burner.

**Platform evaluation**

To evaluate the potential of our recyclable PCBs for real-world applications, we create a functional prototype of an IoT sensor to monitor environmental conditions such as temperature, humidity, and pressure. Such sensors provide critical information for environmental and smart-building monitoring, and this industry is projected to grow by billions of new devices in the coming years which raises a pressing need to mitigate the environmental harms of the e-waste they will generate[36]. In addition to being a high-impact application, a complete IoT sensor requires meeting numerous technical specifications which highlight the versatility of our vPCBs. The platform must perform all the basic functions of a small computer, demonstrating the potential for using vPCBs in a variety of computing devices. Beyond the standard requirements of consumer electronics, IoT devices must also be capable of high-frequency signaling for transmitting and receiving radio signals at GHz frequencies which are affected by PCB properties such as dielectric constant and loss tangent.

We create an end-to-end prototype of a wireless environmental sensor with vPCB according to the schematic shown in Fig. 4a and deploy it on a server rack as shown in Fig. 4b. Our device includes a microcontroller with an integrated Bluetooth radio, a digital sensor, a coin cell battery, onboard chip antenna, and assorted passive components (see Methods for additional details). We fabricate the circuit on both vPCB and standard FR-4, and program both to wirelessly transmit sensor measurements from a data center setting for 16 hours. Figs. 4c-e show the resulting temperature, pressure, and humidity. Both devices were able to successfully transmit Bluetooth packets and no gaps are observed in the data from device failure. The data show the sensors themselves are closely correlated due to their proximity, but with small fixed offsets likely due to individual sensor variation or their small physical separation.

To further qualitatively analyze the signal integrity of circuits on vPCB, we perform eye diagram measurements that reflect vital parameters for signal in digital transmissions, such as signal-to-noise ratio (SNR) and clock timing jitter. A well-routed signal line is designed and patterned onto both vPBC and FR-4. A serial BERT (Keysight N4903A) is used to generate a 2.48 Gb/s pseudorandom binary sequence (PRBS), and the eye diagram is measured by a wide-bandwidth oscilloscope (Keysight 86100D). Our vPCB has nearly identical performance as standard FR-4 at a glance (Fig. 4f).

**Fig. 4: Platform evaluation**

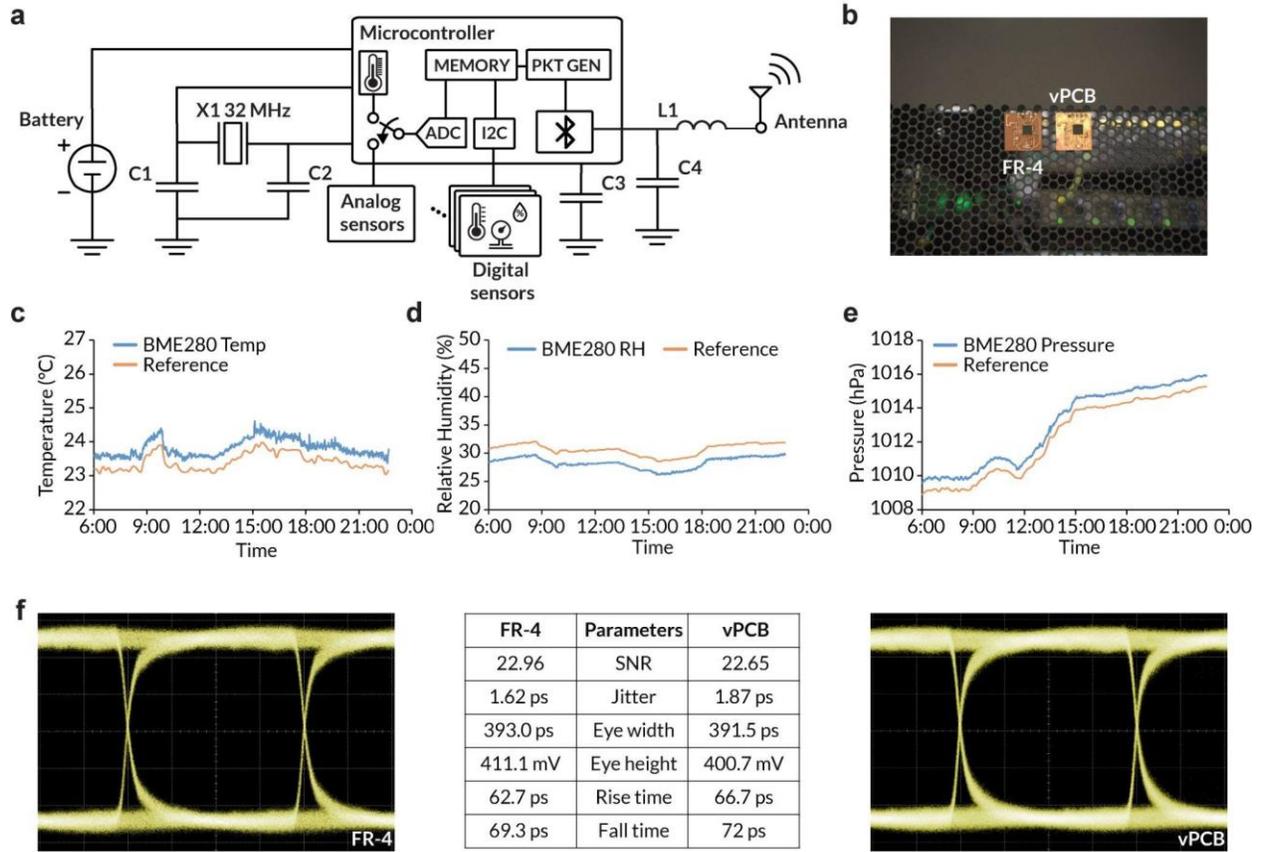

**a,** Circuit block diagram. ADC, analogue-to-digital converter; PKT GEN, packet generation; C1–C2, capacitor 1–capacitor 2; L1, inductor 1; X1, 32 MHz crystal oscillator 1; I2C, Inter-Integrated Circuit. **b,** Photograph of experimental set-up. **c, d, e,** Data center temperature (**c**), relative humidity (**d**), and pressure (**e**) measurements from two platforms over 16 h showing successful continuous running and high-frequency signal transmission. **f,** Comparison of eye diagrams of FR-4 and vPCB.

**Repair and remanufacturing**

A unique property of our vPCBs is that the vitrimer's dynamic covalent bonds enable reliable healing of physical damage such as fractures and holes. This opens up the ability to repair and remanufacture PCBs to reduce their environmental impact and promote a circular economy. A recent study from Microsoft[37] showed that repairing broken devices with replaceable parts resulted in a reduction of up to 85% in greenhouse gas (GHG) emissions and 90% in waste avoidance compared to replacing the whole device. Once these devices are beyond repair, we explore the potential of vPCBs to enable value cycling by repairing physically damaged base materials and remanufacturing them into new ones.

A repair method is developed that heals holes and fractures as shown in Fig. 5a. Copper is first removed through chemical etching. Subsequently, a small amount of fresh vitrimer material in liquid form is applied to fill the holes and follow the same heat press process shown in Fig. 2b to create a new CCL. The repaired GFRV is shown in Fig. 5a, wherein all the through holes have been healed with no discernable marks indicating the hole location. The joint strength of healed via holes is also evaluated using a shear punch test (see Methods for details and Extended Data Fig. 3). A comparison is made between the FR4 with repaired via holes using cyanoacrylate glue and the results indicate that vitrimer creates a stronger interface at the hole boundary, avoiding catastrophic failure as seen with the superglue bond that breaks. The healing capability of vitrimer chemistry enables the creation of new covalent bonds between fresh vitrimer and GFRV. Specifically, the transesterification reaction occurring in the vitrimer system leads to topological rearrangements while preserving the integrity of the molecular network[26,38]. This healing mechanism cannot be achieved by those of traditional thermoplastics or thermosetting plastics.

Using the same fabrication steps described in the previous section we can use the resulting bare GFRV to create new circuits.

To evaluate our healed composite, the copper adhesion is first measured after four cycles of copper re-lamination as shown in Fig. 5b. Interestingly, we find that the interfacial adhesion improves after each repair cycle. This is likely attributed to the micro-scale indentations on the GFRV surface caused by heat-pressed copper foil, as observed using SEM shown in Fig. 5c. These indentations increase the surface roughness after each remanufacturing attempt, which is known to improve adhesion[39]. The dielectric constant and volume resistivity of the repaired material are evaluated following and found to remain within the range of common FR-4 varying by a maximum of 6.5% even after four repair cycles as shown in Fig. 5d,e.

vPCBs also have shape-memory properties that can be used to reverse physical damage that can occur due to mechanical stress such as GPU sagging. Mechanical deformation of the vPCB below its $T_v$ results in a temporary shape change. Subsequent heat treatment at 100 °C (above both Tg and Tv) for 1 min prompts the material to revert back to its original, pre-deformation configuration. Figure 5f demonstrates a GFRV that has been deformed under an external force, then successfully recovers to its thermodynamically favored original shape after heat treatment. This shape-memory attribute not only underscores the inherent self-healing properties of vitrimers but also further highlights the potential for the circular reuse of mechanically damaged vPCB materials.

**Fig. 5: Repair and remanufacturing of vPCB**

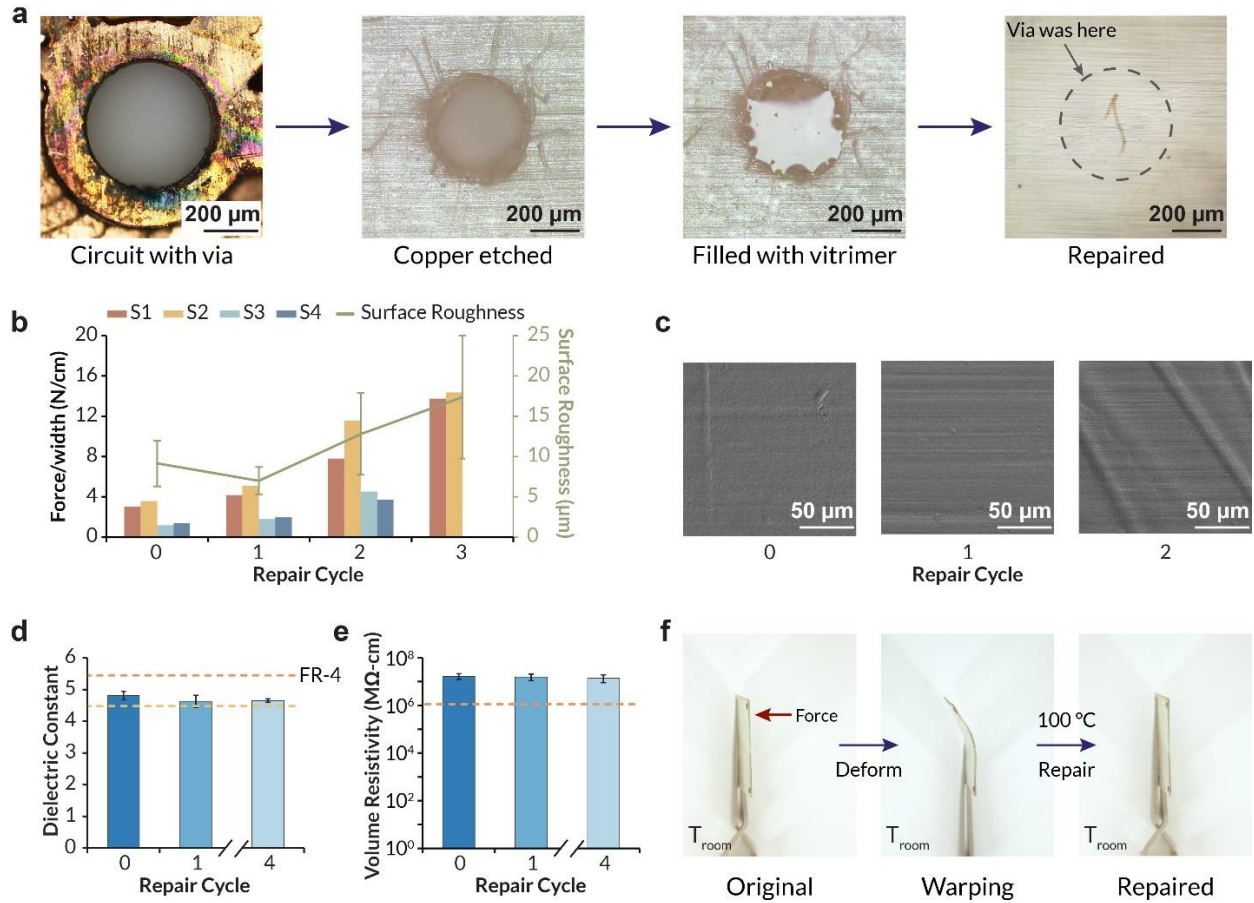

**a,** Optical micrographs of a via on vPCB being healed. **b,** Average peeling force of copper-clad after three repair cycles for various specimens with secondary vertical axes showing the corresponding surface roughness before each repair cycle ($N = 4$, error bars $= \pm\sigma$). **c,** SEM images of GFRV surface after each repair cycle, showing orderly indentations caused by heat-pressed copper foil. **d, e,** Characterized dielectric constants (**d**), and volume resistivities (**e**) of repaired GFRV after each repair cycle ($N \geq 3$ (**d**) and $N = 1000$ (**r**), error bars $= \pm\sigma$). Remanufactured vPCB exhibits nearly identical dielectric constants and volume resistivities. **f,** Thermadapt shape-memory behavior of GFRV. GFRV is deformed under external force, and is recovered to its original shape after being heated at 100 °C for 1 min. This allows for the repair of warped vPCBs at elevated temperatures, making it a more environmentally friendly option compared to traditional PCBs.

**Closed-loop recycling**

Vitrimer-based PCBs can be repeatedly repaired through the process described above, however these PCBs will eventually reach the end-of-life when a design becomes outdated or if repair is no longer economically feasible. To maximize circularity, we further develop a closed-loop recycling process for end-of-life vPCBs to nondestructively disassemble the fiber composites back into high-quality raw materials for use in creating new, fully functional circuit boards.

We begin the vPCB recycling processes by removing the components, cleaning the surface, and then dissolving the copper using ferric chloride to isolate the GFRV substrate (see Methods for details). Fig. 6a illustrates our process for deconstructing the GFRV into raw materials. We immerse the GFRV sample in various polar aprotic solvents for the extraction of glass fibers. We examined acetone, chloroform, dimethylformamide (DMF), and tetrahydrofuran (THF). We found that DMF and THF do not react with the GFRV materials but can ingress into the polymeric network due to their small molecules, resulting in swelling (Extended Data Fig. 4, Supplementary Video 1). This swelling enables the separation of the vitrimer matrix and glass fibers without chemical degradation. We select THF for our recycling process, as it achieves similar swelling performance but has a low boiling point of 66 °C and enables easy solvent removal to recover the vitrimer. In contrast, DMF has a higher boiling point of 153 °C and requires rotary evaporation which may result in the loss of the catalyst in the material (see Supplementary 1.1 for additional discussion).

The vitrimer matrix is fully separated from the glass fiber layers after immersion in THF for 96 hours (shown in Fig. 6a, details are provided in the Methods). The recovered vitrimer and glass fiber are shown in Fig. 6a, b, respectively, and can be reused directly after drying. Vitrimer polymer can be easily recovered with ~98% isolated yield after being precipitated out from the THF, and 100% of the glass fiber is recovered. We also demonstrate a THF recovery efficiency of approximately 91%, with 1.5% being lost due to evaporation during testing and measurements, and 7.5% remaining in the swollen vitrimer matrix which was not attempted to recapture. The THF recovery efficiency could be further optimized to over 97% by performing recycling in a properly designed and operated recovery system[40]. This is a marked improvement over previous solvolysis methods, which necessitate additional adsorption or distillation procedures to remove the solvent and recover the polymer matrix.

We create new GFRV samples using recycled materials by grinding the recovered vitrimer into a fine powder, mixing it with a 40 wt% fresh vitrimer matrix in viscous liquid form to fully penetrate the small interspaces in our dense glass fiber weave, and applying the same heat press process described above. We note the required pressure has an inverse quartic relationship to pore size according to the Hagen Poiseuille equation suggesting the potential for using 100% recycled vitrimer by adjusting the material or process parameters (see Supplementary 1.2 for additional discussion). The resulting reformed GFRV is shown next to a newly made GFRV sample in Fig. 6c. The resulting composite showed no apparent differences except for slight discoloration on the surface due to thermal oxidation.

We hypothesize above that swelling does not degrade the materials; however, this method raises a number of fundamental questions of whether the process removes non-covalently bonded catalyst critical to the transesterification vitrimer's properties, whether it affects the crosslink density, or changes the polymer backbone. Next, we perform a series of experiments to investigate our hypothesis that small molecule solvent-based swelling does not degrade the materials and understand the underlying mechanisms of our recycling process. We first examine SEM images of reformed GFRV which show smooth surfaces without any noticeable damage in the form of microcracks or pores (Fig. 6c). At a macro-scale, this confirms that the dynamic nature of the polymeric macromolecular network is preserved after every recycling as the polymer can flow and form a continuous sheet from pulverized recycled vitrimer powder without being dissociated by the solvent.

The mechanical behavior of the recycled vitrimer matrix is evaluated by comparing the storage modulus and tan delta values post-recycling (Extended Data Fig. 5a,b). We quantify the retention of storage modulus post-recycling, observing 96.9% after one recycle cycle and 94.4% after two cycles (Extended Data Fig. 5c). Although we observe some narrowing and broadening of the tan delta peak between cycles (see Supplementary 1.3 for additional discussion), the high storage modulus retention clearly attests that the reduction in crosslink density induced by our swelling recycling approach is negligible. To evaluate the potential loss of non-covalently attached catalyst during our recycling procedure, we leveraged Maxwell's relation for viscoelastic materials to derive Arrhenius curves[41] (Extended Data Fig. 5d). We determined a Tv of 79.6 °C for pristine vitrimer. Upon recycling once, the Tv was slightly reduced to 78.3 °C and 79.4 after recycling twice. The marginal shift in Tv implies no loss of the non-covalently bonded catalyst throughout our THF swelling recycling process across multiple cycles, as a reduction in catalyst concentration within vitrimer would precipitate a substantial increase in Tv[41].

To further confirm the maintained chemical composition of the vitrimer polymer backbone after the recycling process, samples at different recycling stages are analyzed using Fourier-transform infrared (FTIR) spectroscopy and compared with a new GFRV sample in Fig. 6e. We observe the disappearance of 903 cm[-1] for epoxy functional group after the vitrimer is cured, and the appearance of sharp peaks at 1037, 1726 and broad intense peaks at 3359 cm[-1] for -CO-O-, -C=O, and -OH groups respectively. This indicates the consumption of all epoxy end groups for the formation of ester linkages with crosslinked networks in vitrimers. The recovered vitrimer retained identical functional groups with characteristic peaks as the original, as shown in Fig. 6d, which indicates that recycled vitrimers maintain their crosslink density and undergo exchange reactions without changing the chemical polymer backbone. Figure 6e also indicates that the vitrimer can resist ferric chloride corrosion, which enables the production of vPBC through the conventional and scalable PCB manufacturing technique of chemical etching. The properties of GFRV made of recycled vitrimer are also highly comparable to the original one (Extended Data Fig. 6).

## Fig. 6: Recycling of vPCB

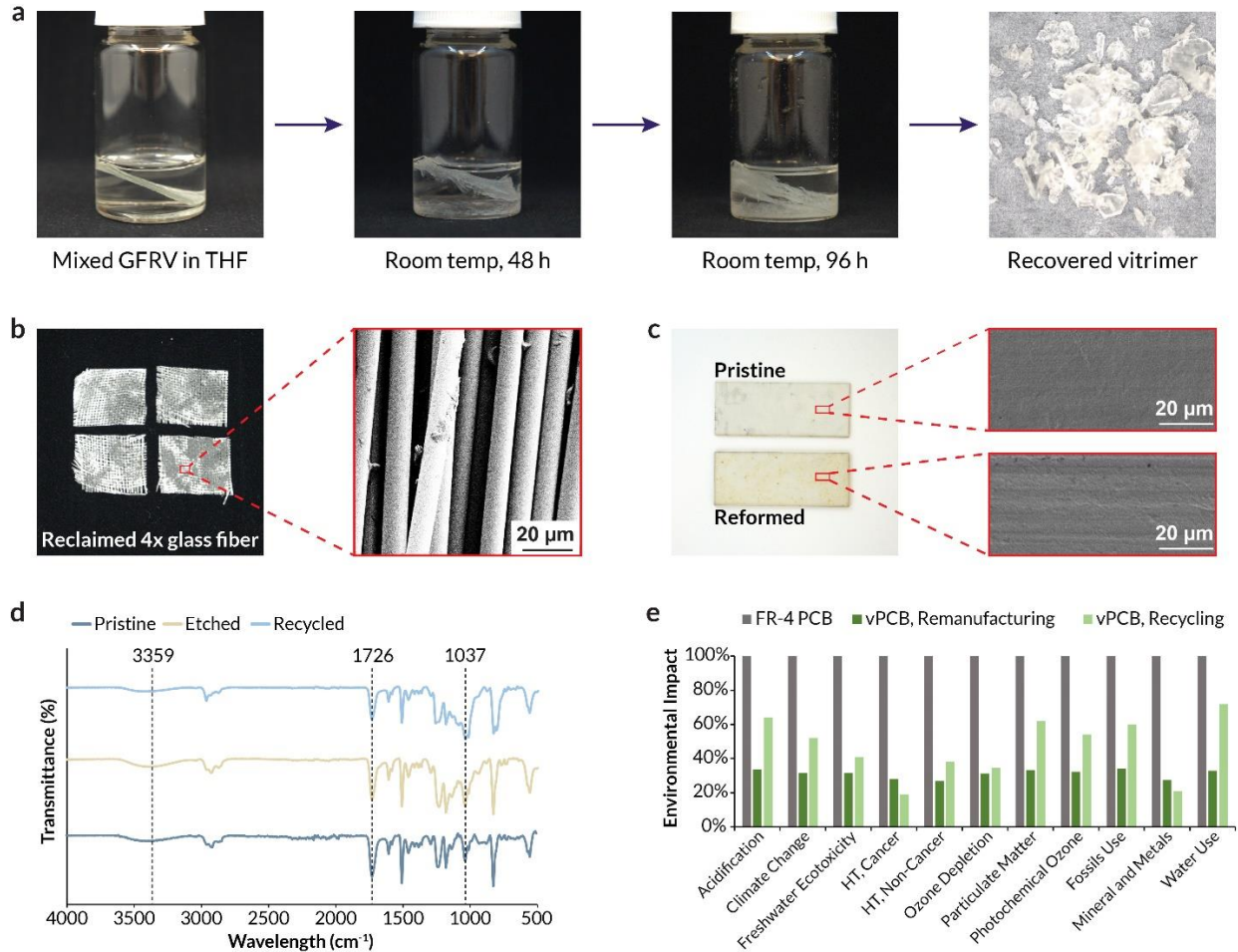

**a,** Photographs showing recycling procedures of GFRV from obsolete PCB and recovered vitrimer matrix after precipitation from the THF. **b,** Photograph of reclaimed glass fiber weaves and SEM image showing the clean microstructure from recycling. **c,** The recovered vitrimer was reused to form new GFRVs. The appearance of reformed composite maintains nearly identical to the pristine, and SEM images show smooth surfaces of reformed GFRV without any microcracks or pores. **d,** Comparison of FTIR spectra of pristine vitrimer, vitrimer after etching, and recovered vitrimer from THF. **e,** Environmental impact comparison of conventional FR-4 PCB and vPCB in end-of-life disposal scenarios across 11 different categories. vPCB is assumed either four cycles of remanufacturing or recycling, with a THF recovery efficiency of 97% in the recycling scenario. As a comparison baseline, FR-4 PCB is assumed to have 5 lifetimes with the best-case assumption for end-of-life incineration. The vPCB results are normalized to the FR-4 PCB for each category respectively.

**Life-cycle assessment**

When proposing new remanufacturing or recycling processes, it is important to assess their environmental impacts or benefits using tools such as LCA[42]. We perform an LCA to evaluate the environmental impact of our vPCB and compare our remanufacturing and recycling processes to conventional PCB fabrication and disposal at scale. Our cradle-to-cradle LCA uses industrial models and encompasses material synthesis, CCL manufacturing, freight transportation, and end-of-life PCB disposal or recycling.

Notably, the GFRV fabrication process produces *cured* laminates. This is a key advantage over conventional prepreg materials which are only partially cured and have a limited shelf life of 3-6 months[32]. Maximizing the lifetime of conventional prepreg materials requires refrigeration and low-humidity environments, thereby increasing their environmental footprint. To quantify this difference, we compared the environmental impact of vPCB freight with conventional FR-4 prepregs. The LCA results demonstrate a substantial reduction of 20.0% in global warming potential, 28.0% in ozone depletion potential, 14.0% in fossil depletion, and 26.0% in mineral and metal use (Extended Data Fig. 7).

In order to compare the environmental impact of vPCBs with conventional PCBs in end-of-life scenarios, we use the business-as-usual standard of incineration as the baseline for the latter. Our model makes the best-case assumption for conventional PCBs, in which 89.3% of the waste heat and 11.1% of the waste electricity will be recovered and credited as energy production. Our remanufacturing scenario models the case where a product is repaired and remanufactured, with certain parts replaced as necessary. Our LCA model looks specifically at the environmental cost of remanufacturing the PCB using the method described in the section above. In the recycling scenario, the product is broken down into its component raw materials, which are then reused in creating new PCBs.

The results, as shown in Fig. 6e, indicate that vPCBs have the potential to substantially reduce numerous environmental impacts including acidification, global warming potential, and ozone depletion potential. Extended Data Fig. 8 shows a detailed breakdown of the carbon footprint of a conventional PCB indicating that raw materials make up 48.5% of the total environmental cost. We show however that leveraging the unique healing properties of our vPCB enables remanufacturing with significantly less material and energy and allows this process to be repeated for multiple cycles to further reduce impact. Our LCA assessment indicates that this method would reduce all of the studied environmental impact metrics by over 65% with four remanufacturing cycles.

The primary benefits of using vPCBs are reducing end-of-life byproducts, disassembling end-of-life vPCBs into raw materials and creating new, fully functional circuit boards. The environmental impact resulting from four recycling cycles shows a remarkable reduction in 11 metrics of up to 80.9%. This includes a 35.8% reduction in acidification, 47.9% in global warming potential impact, 59.1% in freshwater ecotoxicity, 61.8% in human non-cancer toxicity emissions, 65.5% in ozone depletion potential, 38.0% in particulate matter, 45.9% in photochemical ozone, 40.2% in fossils depletion, 79.2% in mineral and metals use, and a 28.1% reduction in water use over four recycling cycles. It is noteworthy that the use of vPCBs could lead to savings of up to 80.9% in human cancer toxicity emissions.

In this study, we have reported a recyclable PCB based on transesterification vitrimers and a nondestructive, swelling-based separation for vitrimer composite recycling. Our design achieves the following three key goals: 1) raw materials with strong adhesion between the vitrimer material and copper layers on par with FR-4 PCBs, 2) compatibility of vPCBs with the traditional, inexpensive PCB manufacturing processes to enable scalable production including multi-layer designs, 3) electrical and mechanical properties comparable to standard FR-4 PCBs required to create functional circuits. Such recyclable PCBs with excellent electrical and mechanical properties, chemical etchant resistance, repairability, and closed-loop recyclability, make them a promising and straightforward alternative solution to conventional PCBs. Not only do they reduce the toxic byproduct problems associated with conventional PCBs, but they also significantly improve a broad range of environmental impact metrics. Additionally, the abundance and widespread commercial availability of epoxies and catalysts with varying properties create numerous opportunities to experiment with new vitrimer formulations optimized for specific properties. Our approach of swelling PCBs could also be explored for automated disassembly of circuit components to simultaneously enhance value recovery from ICs while recycling raw materials. We hope this study will open up new directions for the development and application of emerging recyclable polymeric materials and composites for the electronics and computer industry that centers on environmental sustainability.

## Data availability

All data needed to evaluate the conclusions of the paper are available in the paper or in the Extended Data and Source Data.

## Acknowledgments

We thank T. Cheng for discussion, Z. Englhardt for help with Bluetooth coding, B. Kuykendall for the use of mechanical testers, C. Li for feedback on the figures, K. Liao and M.Parker for help with flammability testing, and H. Wang for help with composite fabrication. We also thank D. Baker, F. Newman, and C. Toskey for help with sputter coating and copper plating. This research is funded by the Microsoft Climate Research Initiative.

## Author Information

Authors and Affiliations

**Paul G. Allen School of Computer Science & Engineering, University of Washington, Seattle, WA, USA**

Zhihan Zhang, Jake A. Smith, Bichlien H. Nguyen, Shwetak Patel & Vikram Iyer

**Department of Mechanical Engineering, University of Washington, Seattle, WA, USA**

Agni K. Biswal, Ankush Nandi & Aniruddh Vashisth

**Microsoft Research, Redmond, WA, USA**

Kali Frost, Jake A. Smith & Bichlien H. Nguyen

Corresponding author

Correspondence to Aniruddh Vashisth or Vikram Iyer.

Contributions

Z.Z., B.H.N., A.V. and V.I. conceptualized, organized and structured the work. Z.Z., A.K.B. and A.N. fabricated glass fiber reinforced vitrimer composites. Z.Z. manufactured vitrimer-based PCB and conducted characterizations. Z.Z. designed the hardware system, experiments and evaluations. Z.Z., J.A.S., and B.H.N. designed the repair experiments and evaluations. Z.Z., A.K.B., J.A.S., B.H.N. and A.V. designed the recycling experiments and evaluations. Z.Z. and A.K.B conducted material characterizations. K.F. conducted the life-cycle assessment analysis. Z.Z. and V.I. wrote the manuscript. All authors contributed to the study concept and experimental methods, discussed the results and edited the manuscript.

## Ethics declarations

Competing interests

K.F., J.A.S., and B.H.N. are employees of Microsoft Corporation. S.P. is a Google employee. Z.Z., A.K.B., A.N., A.V. and V.I. declare no competing interests.

# Extended data figures and tables

Extended Data Fig. 1 Dynamic mechanical analysis of pristine vitrimer and recycled vitrimer

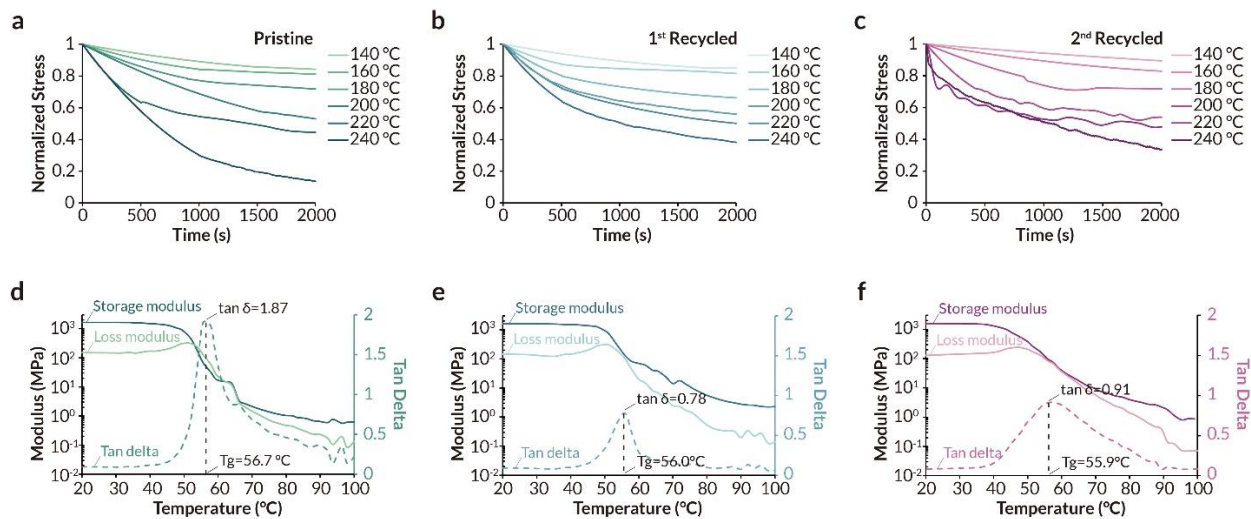

**a, b, c,** Normalized stress relaxation curves of pristine vitrimer (**a**), vitrimer after one recycling cycle (**b**), and vitrimer after two recycling cycles (**c**) at temperatures ranging from 140 °C to 240 °C. In all cases increasing temperature results in faster stress relaxation. **d, e, f,** Characterized storage modulus, loss modulus, and tan delta of pristine vitrimer (**d**), vitrimer after one recycling cycle (**e**), and vitrimer after two recycling cycles (**f**).

Extended Data Fig. 2 Peel strength for laminates with a layer of partially cured vitrimer

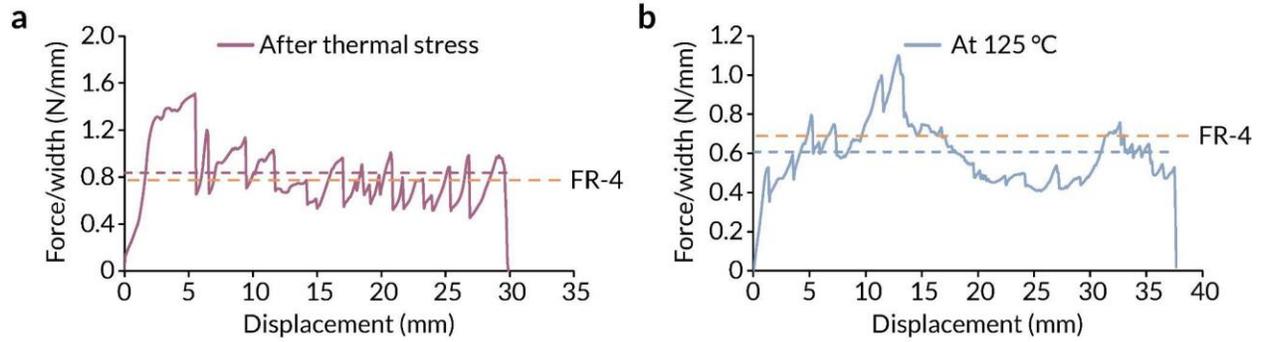

**a, b,** Curves of the peeling force per width of copper-clad versus displacement for laminates with a layer of partially cured vitrimer after thermal stress (**a**), and at 125 °C (**b**) compared to the PCB standard of FR-4.

Extended Data Fig. 3 Joint strength of repaired via holes in GFRV

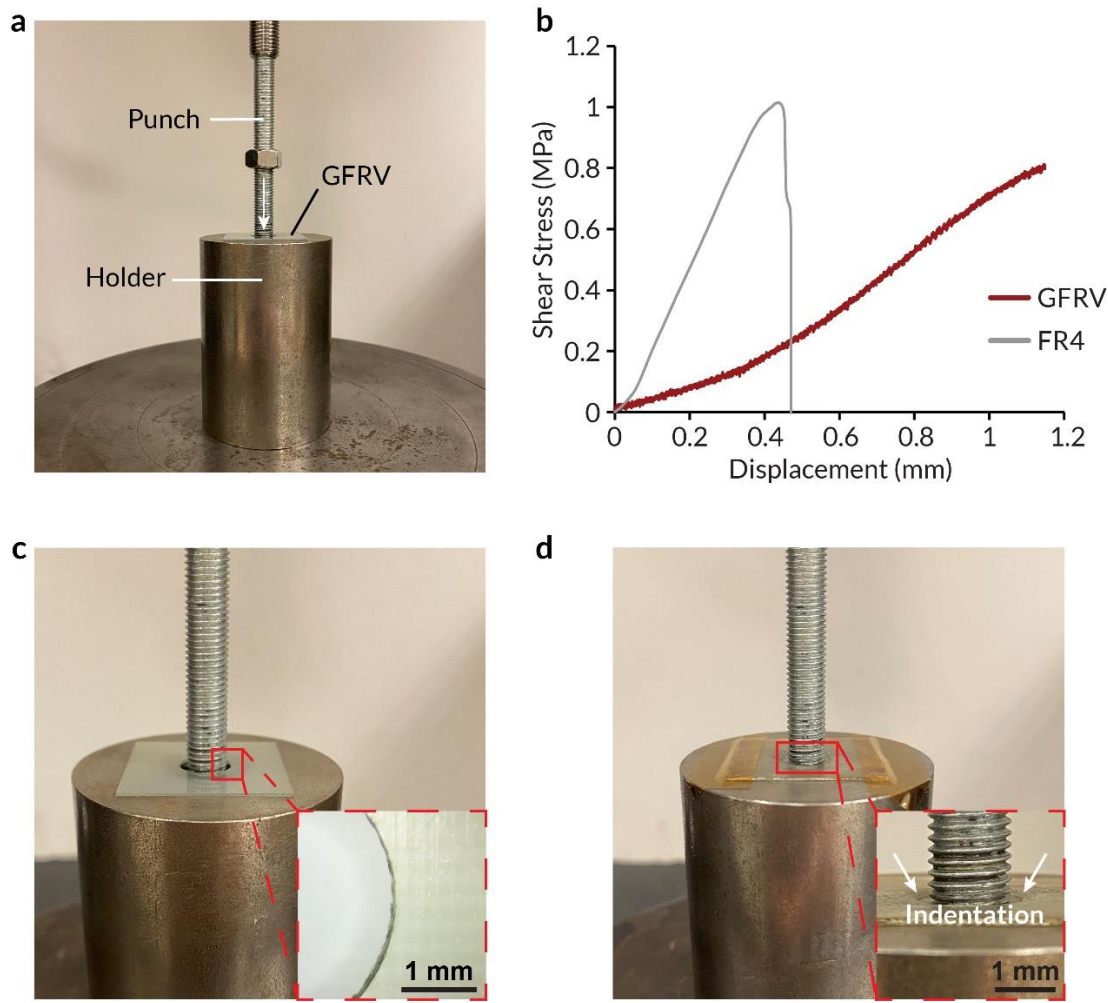

**a,** Photograph showing the joint strength testing setup. Specimen is centered on a metal hollow cylinder support with a support span of 16 mm. **b,** Characterized shear stress of repaired via holes in GFRV compared to the repaired holes in FR-4 using super glue. **c,** Photograph of FR-4 after shear punch, showing cyanoacrylate glue bond broke. **d,** Photograph of GFRV after shear punch, showing the repaired via hole was deformed into a funnel-shape under the force of punch but remained intact, indicating a stronger interface at the hole boundary.

Extended Data Fig. 4 Solvents test for GFRV recycling

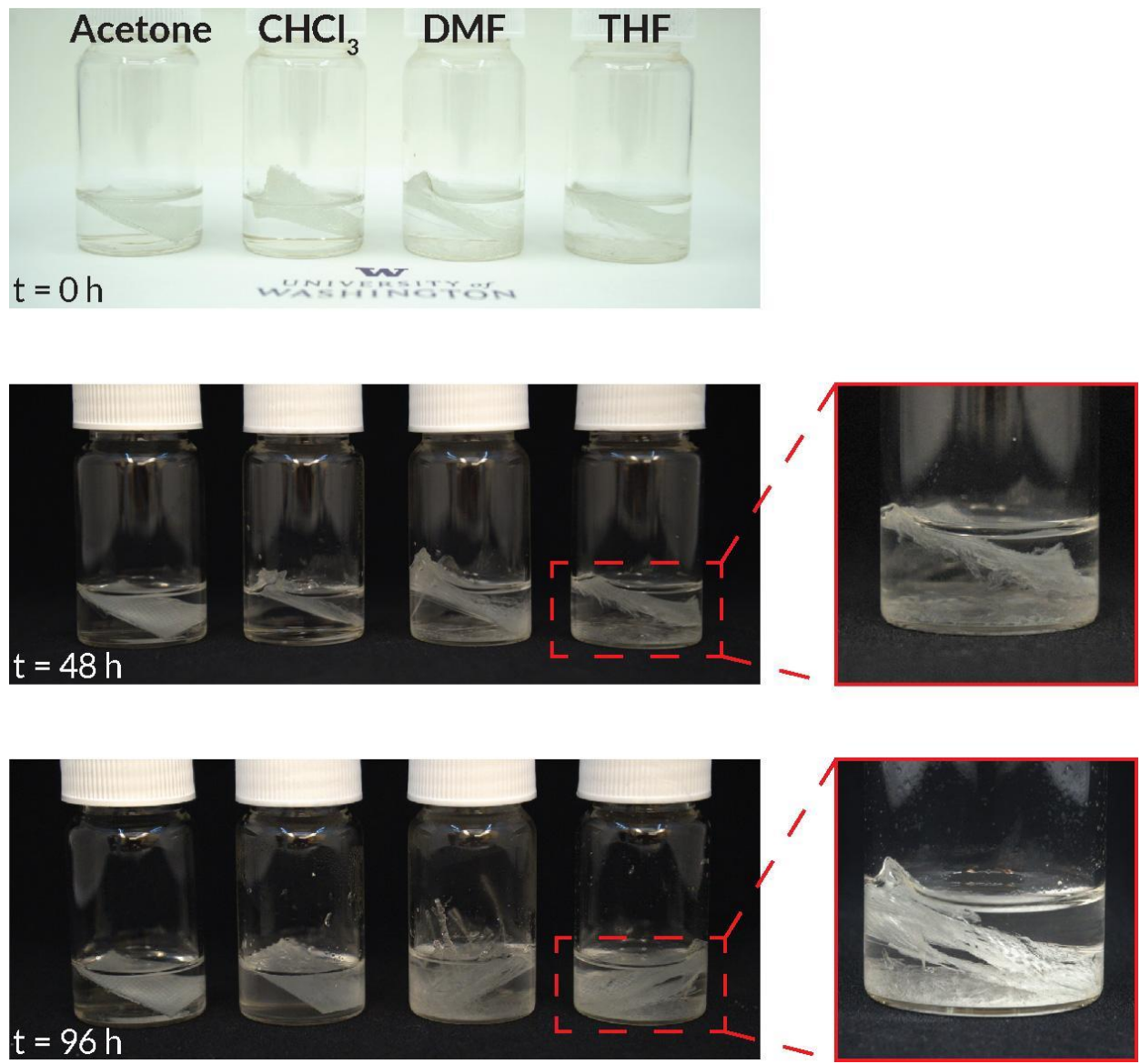

GFRV samples were cut into rectangular shapes and immersed in various solutions (Acetone, CHCl₃, DMF, THF); the top, middle and bottom photos were taken immediately after immersing, after 48 hours, and after 96 hours, respectively.

Extended Data Fig. 5 Characterized storage modulus, tan delta, retention of storage modulus, and vitrimer transition temperature of recycled vitrimer

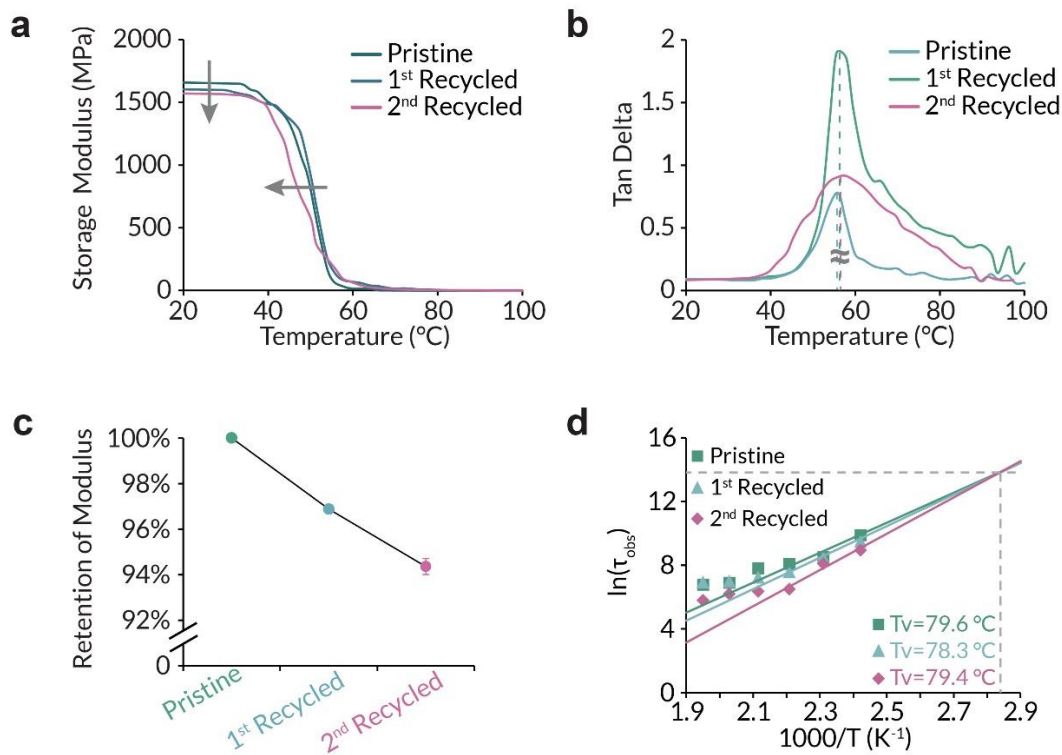

**a**, Characterized storage modulus temperature sweep results of vitrimer after one and two recycling cycles compared to pristine. The storage modulus shows a slight decrease after recycling. **b**, Tan delta temperature sweep results of vitrimer after one and two recycling cycles compared to pristine, tan delta broadens and the left shift of peaks is negligible after recycling. **c**, Retention of storage modulus ($N = 3$, error bars = $\pm\sigma$) of vitrimer after one and two recycling cycles compared to pristine. **d,** Tv comparison of pristine vitrimer, vitrimer after one and two recycling cycles, indicating the shift of Tv is negligible after recycling. The Arrhenius plot is derived with a linear fit to the low-temperature region (140 °C to 180 °C), and its intersection with where the stress-relaxation constant is 10^6 indicates the Tv.

Extended Data Fig. 6 Characterized electrical and mechanical properties of reformed GFRV

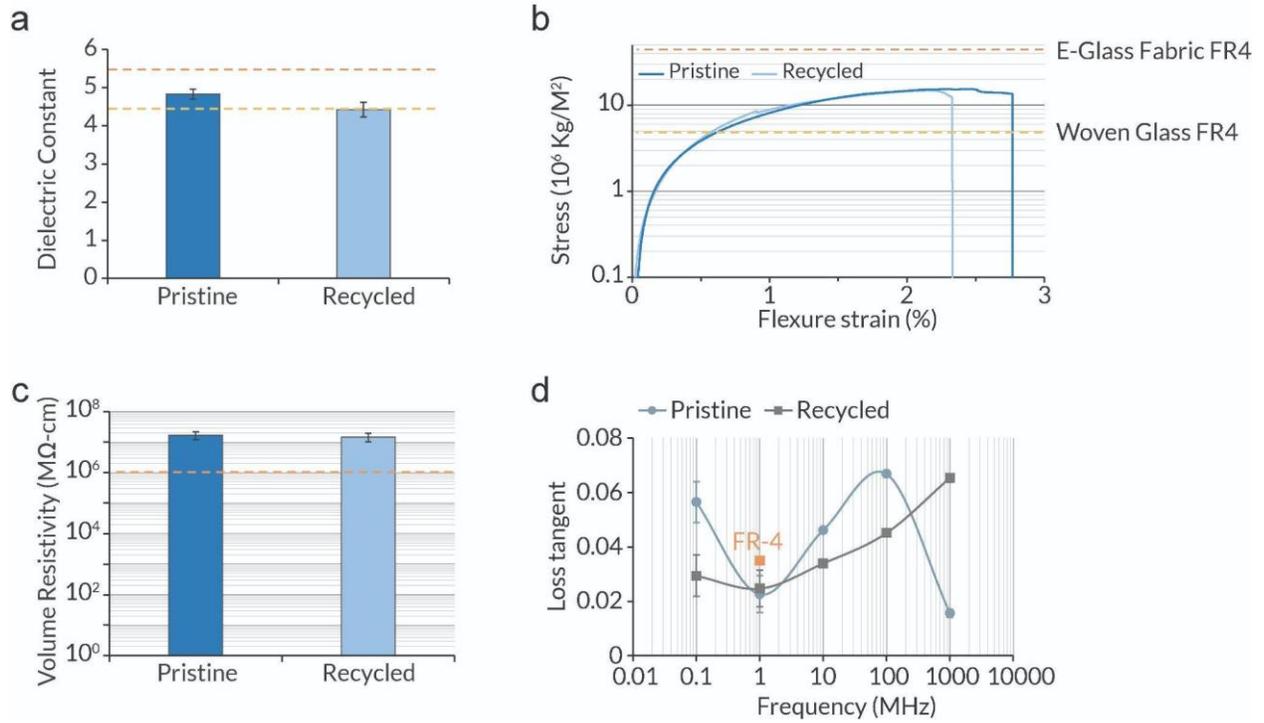

**a, b, c,** Characterized dielectric constant (**a**), flexural strength (**b**), volume resistivity (**c**), and loss tangent (**d**) of reformed GFRV compared to virgin composite ($N \geq 3$ (**a, c, d**) and $N = 1000$ (**c**), error bars $= \pm\sigma$).

Extended Data Fig. 7 Environmental impact of vPCB freight

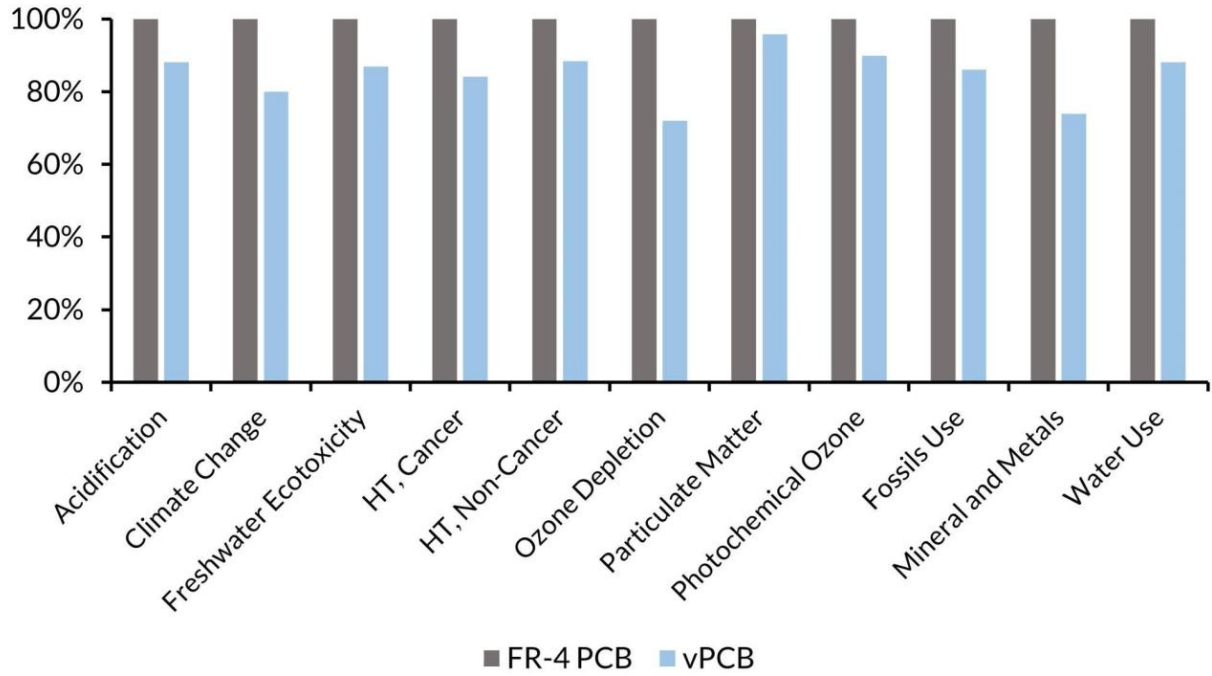

Comparison of the environmental impact of vPCB freight versus conventional FR-4 prepreg freight across 11 different categories.

Extended Data Fig. 8 Global warming potential impact breakdown of conventional FR-4 PCB

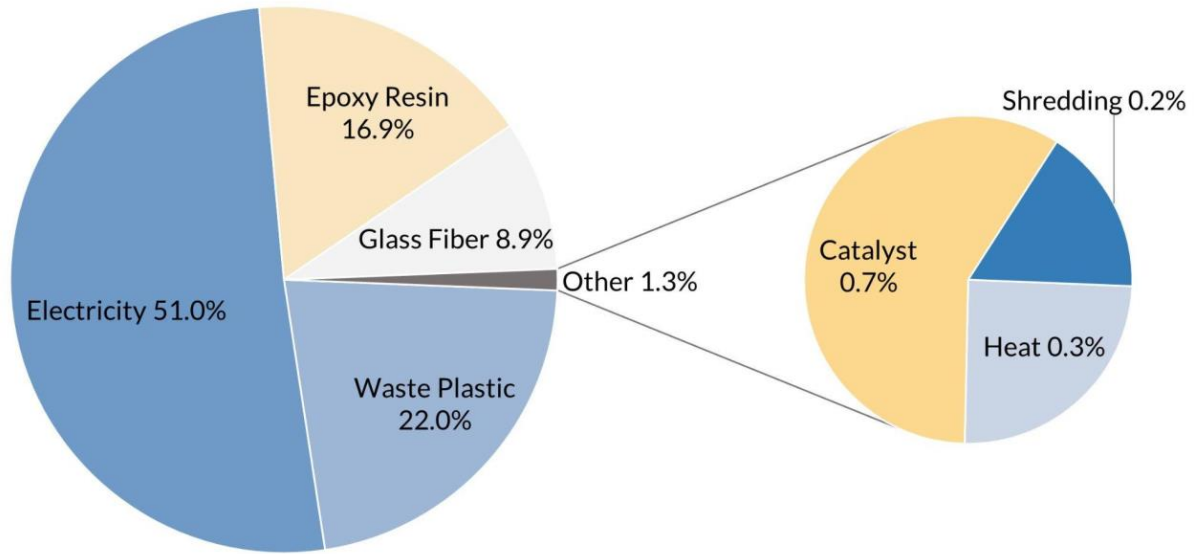

Detailed contributions in kg $CO_2$ equivalents to global warming potential for conventional FR-4 PCB.